# A Decentralized IoT Data Marketplace

Pooja Gupta
Computer Science & Engineering
UNSW
Sydney, Australia
pooja.gupta@student.unsw.edu.au

Salil S. Kanhere
Computer Science & Engineering
UNSW
Sydney, Australia
salil.kanhere@unsw.edu.au

Raja Jurdak
Data61, CSIRO
Queensland, Australia
Raja.Jurdak@data61.csiro.au

*Abstract*—This paper proposes an architecture for dynamic decentralized marketplace for trading of Internet of Things (IoT) data. To this end, we introduce a 3-tier framework which consists of provider, consumer and broker. The framework is realized using multiple trustless broker which matches and selects potential data provider based on the consumer's requirements. Rather than using a centralized server to manage the contract between provider and consumer, the framework leverages smart contract-based agreement for automatically enforcing the terms of the contract to the involved parties.

*Keywords—Marketplace, Internet of things, smart contract, Blockchain*

## I. INTRODUCTION

In the coming years, the Internet of Things (IoT) presents an enormous opportunity to transform society by unlocking and unleashing a world of data that, until now, has either been uncollected or has sat largely unused. As the number of data sources expands exponentially, businesses are investigating ways to harness the data for insights they can use, for example, to drive operational efficiencies, to improve their customers' experience, or both. As an example, a fitness tracking app provider may wish to procure air quality data from weather stations deployed all over the city to suggest pollution-free running tracks to its users. Grocery chains may be interested in obtaining aggregated information about food items stored in smart fridges of customers in a local neighborhood to better manage their inventory.

A new business model referred to as data marketplace [1, 2] is emerging whereby data producers can sell their IoT data to interested consumers. However, data generated from IoT devices possess specific characteristics which needs to be considered while designing such a framework. Firstly, IoT data are generated from heterogenous data providers such as individuals, organizations, industry etc. Secondly, the IoT data types and format are diverse in nature like video feed from the CCTV, images from the surveillance cameras or gps data from the mobile device and thirdly, the highest value of IoT data is when it is traded in real-time, and it loses its value otherwise.

A lot of platforms [3-5] have been proposed on this concept which rely on the traditional centralized brokered approach commonly known as client-server models. The highly computational servers and clouds provides all the functionality of the marketplace like contract mechanism, device registry, storage, discovery and search mechanism. All the participants registered to it and publish their device list or query list depending on their role. The server matches and selects the potential provider relevant to the consumer's requirement and facilitates the trading. However, the cloud server model is not suited to the growing needs of IoT devices due to number of reasons. 1) The infrastructure and management cost of such servers are high and will growing number of IoT devices this cost will increase substantially, 2) it is difficult to create a single cloud platform that can support diverse nature of IoT data type/format, 3) server remains bottleneck and a single point of failure that can disrupt the entire network. A decentralized approach based on a Peer-to-Peer (P2P) communication model would solve the above challenges by distributing the computation and storage needs across multiple IoT devices. Such P2P communication however imposes its own set of challenges, foremost being privacy and security.

Blockchain technology which was first introduced by Nakamoto in the Bitcoin cryptocurrency [5] has attracted significant attention in recent years due to its key features including decentralization, immutability and security. Recent research has explored the use of blockchains in the context of IoT data marketplace. An API-based industrial undertaking IOTA [6] intends to provide a decentralized framework for IOT data marketplace. It uses "Coo" as a full node which is utilized to clear out transactions. IOTA network stops working if by any reason, "Coo" is down leading to centralization concerns. Due to this IOTA is not gaining much popularity. The motivation of [7] is similar to ours, namely to provide a marketplace where owners trade their data for either personal or community benefits. In this paper, the authors present an IoT brokered infrastructure with decentralized and open architecture for settlement. The seller charges the buyer based on the count of messages traded. Both the entities monitor the count separately and send it in the transaction to the smart contract which uses it for the settlement purpose. However, it was the first work to use the smart contract in the context of marketplace, they fail to address other key components of the marketplace such as discovery service, contract creation and reliability. In [8], the authors proposed a secure Publish-Subscribe architecture for cyber physical systems based on blockchain which provides confidentiality and reliability of data and anonymity of subscribers. In this architecture, smart contract is employed for all the activities such as setup, publish, subscribe, match, verification and payment in the system. With the expected increase in the number of IoT devices, this approach is not scalable as blockchain is used to store all the activities happening in the system. References [9-10] proposes a complete decentralized and distributed architecture for IoT data marketplace. The approach of both the papers is same that is to use the blockchain as a database for storing the meta-data of the products and distributed storage is used to store product's detailed information. Notice that the main purpose of the blockchain in these two recent works is to serve as a distributed and immutable data registry, whereas the computing capability of the blockchain smart contract was largely wasted.

The current literature is lacking work on using the computing capability to automate the enforcement of terms of the agreement to the involved parties without any centralized intermediator to manage the contract. This paper proposes a decentralized framework that uses multiple data subscription contracts (DSC) and one register contract to achieve distributed and trustworthy contract mechanism for the involved parties. In this framework, each DSC maintains the details of the subscription in a single subscription table for a provider-consumer pair. The DSC provides functions for executing, adding, updating and removing subscriptions. The register contract is used to register and manage information (involved parties, contract address and methods) of the data subscription contract in a contract lookup table.

The rest of the paper is organized as follows. Section II provides a brief overview of smart contract, section III present details of the overall architecture design, roles of entities and interactions among participants. Details of data trading using smart contract are discussed in Section IV followed by conclusion of the paper in section V.

## II. SMART CONTRACT

In order to appreciate our proposed smart contract-based technique for managing the agreement between the involved parties, this section provides a brief overview about Smart Contracts.
Smart contracts [11] are self-executing contracts with the business logic or terms of the agreement being directly written into lines of code. The smart contract is compiled into bytecode and deployed on the blockchain with unique addresses that can be called by any user of the blockchain. Fig. 1 shows the structure of the smart contracts. Smart contract can consist of several functions. So, application binary interface (ABI) is required to specify which function in the contract to invoke.
A transaction specifying the address and ABI triggers the function in the contract and it executes itself according to the coded terms. The state of the smart contract gets stored in the blockchain which can be supervised by the regulators while maintaining the participant's privacy.

## III. IoT MARKETPLACE FRAMEWORK

In this section, we present a system overview and architectural design for data marketplace as shown in Fig. 2

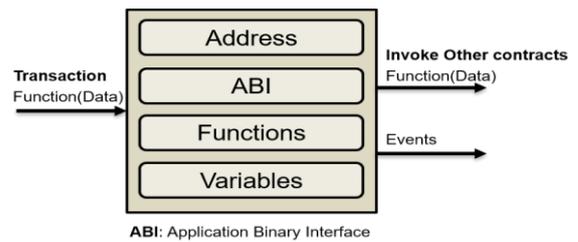

Fig. 1. Structure of Smart Contract.

which consists of multiple participants such as consumers, providers and brokers. Instead of a centralized-trusted-broker, we realize our design using multiple trust-less brokers. As the number of participants in the system increases, it is difficult for a resource-constrained IoT device to maintain direct P2P communication with all the other participants. A highly-resourced broker is thus beneficial in facilitating the data trading by gathering all the data and query lists from the provider and consumer respectively and providing them the potential buyer and seller lists. We incorporate a fee-based mechanism which motivates independent, self-interested brokers to participate in the marketplace. We consider a 3-tier framework in which the first tier consists of IoT data owners, the second tier includes the brokers and the third tier comprises of data consumers. 3-tier framework is comprehended to have defined functionality for the participants of each tier. These nodes are connected by a P2P network. The main roles of the nodes are explained below.

<u>Provider:</u> Providers mainly include the IoT device owners who are interested in selling their data. The owner advertises the data types published by its devices, cost of the data unit in which the owner is interested in selling and specifies the sampling frequency and duration of the data feed. A provider can have multiple heterogenous devices with varying resources and data characteristics. To connect these devices to the network, a conceptual data repository such as Databox [12] can be used. Databox is a protective virtual container which stores data, controls sharing and performs computations over stored data. Even for devices that are not co-located, Databox is capable of aggregating and storing the data of all the connected devices via the Internet. It can also act as the gateway which connects a cluster of local IoT devices to the P2P network.

<u>Intermediate Data Processor (IDP):</u> An IDP can act both as an owner or a consumer depending on its activities in the marketplace. An IDP can buy raw data (acting as a consumer) and use data analytics tools to convert, aggregate and process the data to create value added services which can be sold on the market for profits (acting as an owner). For example, an IDP can bundle data from metro train schedules, real-time information about the location and occupancy of trains to generate taxi demand estimates which could be sold to taxi companies so that taxis are readily available to pick-up passengers. This improves the passenger's satisfaction as they will experience less waiting time and increases taxi companies' profits as more passengers are likely to choose their service.

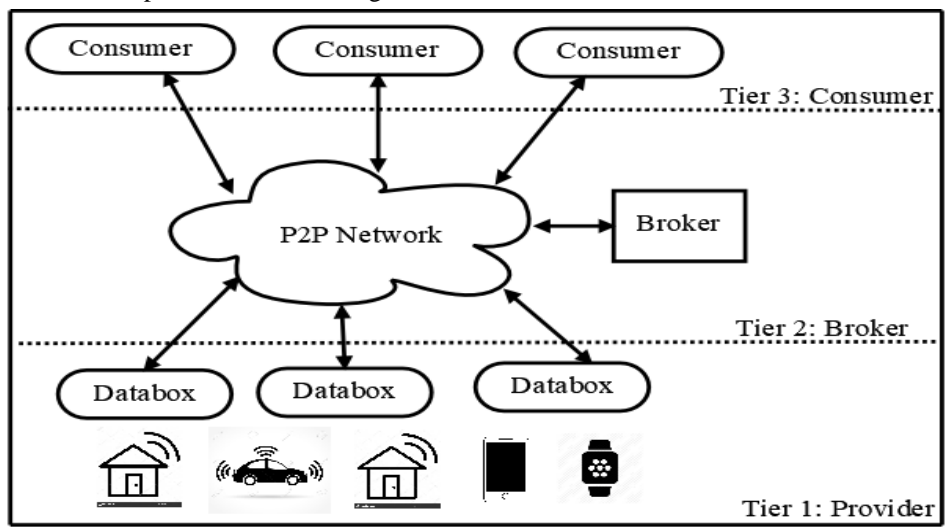

Fig.2. Proposed 3-tier Framework.

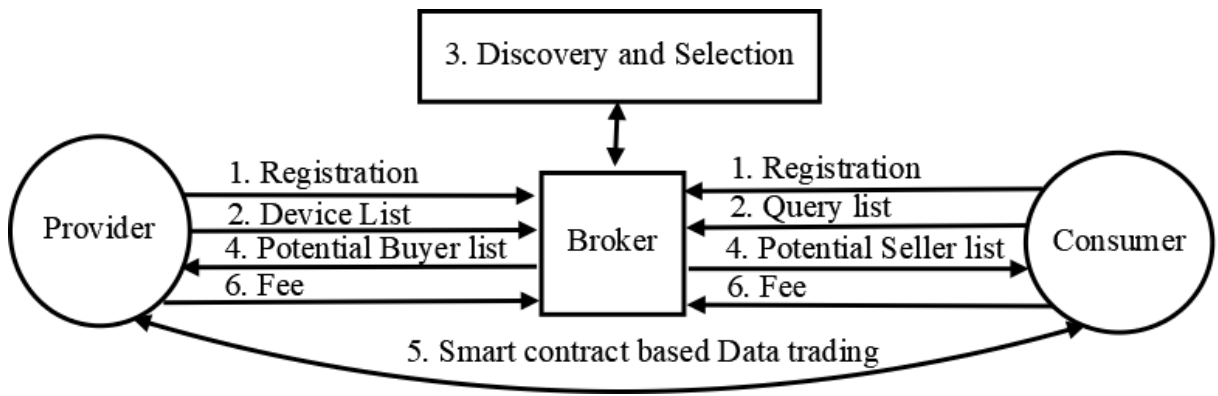

Fig.3. Participants interaction with broker.

Consumer: Consumers are the end users that are interested in buying the raw data or a value-added service (provided by the IDPs). They can be private or government organizations or individual users. A consumer queries the system defining the required data and other aspects. In general, a query list includes data type, data age i.e. archived or real-time data, location, budget, and data frequency. Queries that include location information can support location-based value-added services provided by IDP or limit the search of matching potential seller. For instance, a smart refrigerator manufacturer may need data from refrigerators of their customers that are located in different geographical regions to gain insights into how their devices are being used. The manufacturer will query the marketplace with the intention of collecting extensive usage data about their fridges including temperature settings, number and frequency of door openings, average number of items in the fridge, energy usage, location, etc.

Broker: A Broker is a highly resourced device that will facilitate the trading of data between the consumer and participant. A broker could be a separate entity or virtually created by combining high-resourced participants [13]. Our framework relies on multiple geographically distributed trust less brokers that are interconnected in a P2P network. The main role of the broker is to match the consumer's query with the provider's device list. Since brokers are also trust-less, there is a possibility that they can collude with any participant and provide a biased selected list. To handle this situation, permissioned blockchain is employed which uses smart contract for discovery and search of the potential candidates such as those proposed in [2, 14]. Smart contract usage will eliminate the broker's monopoly market power over how providers are selected and advertised to consumers. All the brokers will be part of this permissioned blockchain and information related to provider's device list are stored in it. The approach is similar to the [10] framework except it would be permissioned and uses smart contract-based discovery and search algorithm. The processing overhead of searching and matching can be handled easily by the resource-rich brokers. The list generated by the smart contract cannot be corrupted by the broker as it is being validated by the other brokers in the network. For each successful data subscription contract, the broker gets a token. The decision to add any new broker is made by the votes of the token-holder. This is an extension of the idea of Token Curated Registry [15]. The token-holders can challenge or accept any new broker to the system to create an efficient and reliable broker network.

The detailed interaction of different participants with broker is shown in Fig 3. All the other participants (i.e. consumers, IDP and providers) must register with a broker that is nearest to their geographical location and create a profile. Since the brokers are synchronized, if a broker fails, then the associated participants are connected to another nearby broker. A consumer can send its query and a provider can send the list of data types it wants to sell to the broker. On receiving the lists, brokers multicast their lists to all the other brokers in the network so that the request can be routed to the appropriate broker based on the specified parameters. The size of such multicast packets and the frequency with which they are sent is subject to the demand. The associated overhead generated will be evaluated in our future work. The broker uses a smart contact-based discovery and selection algorithm and matches the request and demand based on criteria specified by both the parties such as location, data types, budget-pricing etc. It creates the list of potentially matching buyers and sellers and sends it to all the identified participants. It may be possible that there are multiple consumers who want to buy data from a single provider or multiple providers are satisfying the query of a single consumer. A broker sends this one-to-many selected list to the corresponding provider/consumer. The match may involve participants associated with other brokers. For each contract, a broker earns a fee from the involved parties.

However, it could also be possible that no match happens. In this case, broker retains the query and data lists and checks if a match is possible in the future. Since the brokers collectively form a network of trust less nodes where there is no trusted third-party, a trust and reputation framework [8] is employed to rate participants and identify misbehaving participants.

## IV. DATA TRADING

In this section we describe the detailed process of data trading using smart contract. On registration with the broker, the participant becomes a part of the blockchain network and is identified by a changeable public key, which prevents malicious entities from tracking him thus ensuring his privacy. Essentially, each participant has multiple private/public key pairs. The private key is used to sign a transaction while the public key is used to validate the signature of the transaction.

As discussed in section III, after receiving the list of probable providers/consumers from the broker, that consists of the participants' public keys, data types and quoted price, data trading starts as shown in Fig 4. The participant (requester) establishes a direct TCP connection with all the nodes (requestee) mentioned in the list. This off-chain channel is encrypted using transport layer security (TLS).

Two scenarios arise: a) Provider is the requester: Provider compares the budget of all the consumers and proceeds with the negotiation. If there is only one consumer in the list, it accepts the deal and proceeds with the smart-contract. b) Consumer is the requester: Consumer compares the price of the data provided by all the providers and proceeds with the negotiation. If there is only one provider in the list, it accepts the deal and proceeds with the smart-contract. In the negotiation, the requester sends its bid to all the requestees. The requestees respond to the request with a counter bid. The requester evaluates these bids and can either reject and request a new bid or accept the bid.

Since these nodes are part of the blockchain network, they are capable of deploying smart contracts. Setting up a contract requires the nodes to create transactions to run the Application Binary Interface (ABIs) which are functions used to interact with smart contract. Once a bid is accepted by the requester, a smart contract known as data subscription contract is established between both the parties and is compiled and deployed in the blockchain. The data subscription contracts are deployed by the provider for each consumer. For each agreement between a provider-consumer pair, a single contract is created.

A data subscription contract maintains a subscription list, Fig 5, which primarily consists of the data type, contract start time, measurement frequency, subscription period, agreed price and granularity of payment i.e. number of data transactions after which payment needs to be made represented by variable N. The benefit of using N is that if any malicious activity is performed by any of the involved parties, then the settlement is done early instead of waiting till the end of contract. To prevent the malicious behaviour of any participant, both the involved parties deposit an amount, greater than the data charge to account for the broker fee, that is locked for certain time. If a participant is involved in any malicious activity, this deposit is used to penalize it.

The execution of a data subscription contract requires (i) the address where the contract is located and (ii) the ABI. This information is managed by another smart contract named Register contract. The Register contract maintains a contract lookup table as depicted in Fig 6, for recording information regarding smart contracts such as provider and consumer public keys, contract address, and related contract ABIs. The register contract provides two ABIs to fulfil above function. *RegisterContract* receives the information of a new contract and registers its address, associated ABI and other information into the lookup table. The *GetContract* function receives the contract name and returns the address and ABIs of the contract.

The data subscription contract provides multiple ABIs which are used for managing the subscription list and each are designed for specific functions. The provider or consumer can send a transaction indicating which ABI should be run. When a new agreement is established, the provider sends a transaction with ABI *CreateContract* and subscription detail for adding new request for data type in the subscription list and then *RegisterContract* is called to add contract information like contract address and ABIs in the contract lookup table. The deployment of new data subscription is depicted in Fig 7.

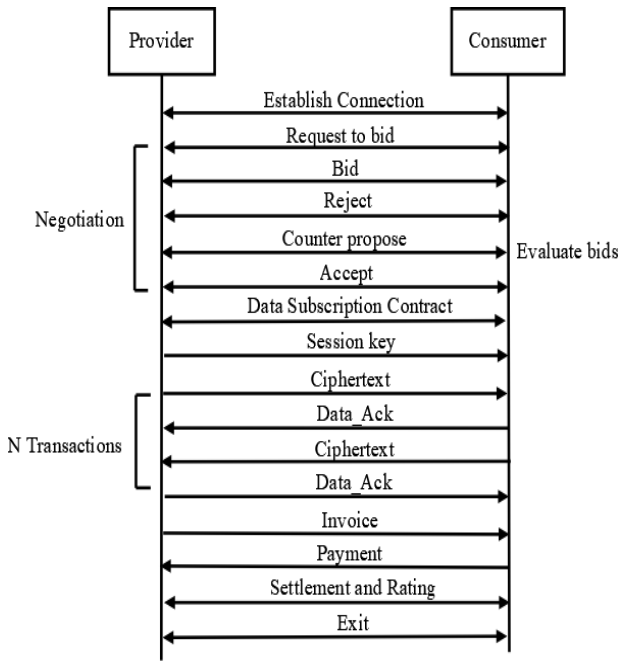

Fig.4. High-level data trading process.

| Device id | Data type | Start Time | Measurement frequency | Session_key | cost | End time | Payment granularity |
|---|---|---|---|---|---|---|---|
| 1 | Temperature | 20/05/2018 00:00 | 30mins | 011001100 001101 | 0.02 | 20/06/2018 00:00 | 100 |

Fig. 5. Subscription Table.

| Contract name | Provider public key | Consumer public key | Contract address | Contract ABI |
|---|---|---|---|---|
| DSC1 | MIGfMA0GCSqCSI b3DQEBAQUAA4G NADCBi.... | GcjDOL5UIsuuuFncZ WBQ7RKNUSesmQR MSGk..... | 0XCA35B7D915458E12ABCDE6968D FE2F44E8FA733C | ExecuteContract(), Settlement()... |

Fig.6. Contract Look-up Table.

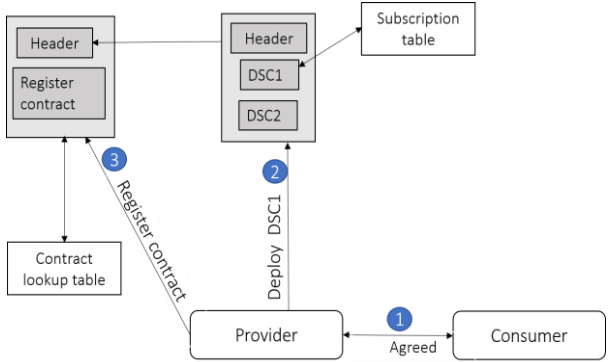

Fig.7. Data Subscription Contract deployment.

The smart-contract based data trading process is illustrated in Fig 8. At the start time of the contract, a transaction is sent by both provider and consumer with ABI *GetContract* to fetch the contract address and ABI. Then using the contract address, the provider calls the ABI *ExecuteContract* which returns Session_key to the consumer and notification mentioning subscription detail to the provider. Session key is the symmetric key used to encrypt all data in transit. Then depending on the agreed contract, data is sent to the consumer and after the consumer verifies the data, an acknowledgement is sent to provider. Both consumer and provider use a metering system to count N transactions. Once N data transfers are done, both the nodes send a transaction with ABI Settlement and counter information. If both the consumer and provider counters match, then an invoice is generated to the consumer and the provider receives the payment. If not, then dispute is lodged and payments are refrained. A reputation system is used to penalize the participants and reduce their rating. This continues until the contract expires. Once the subscription period ends, *RemoveSubscription* is called which allocates a small fee to the broker from the remaining deposit of both the parties. Finally, the subscription is removed from the subscription list at the end of contract.

## V. CONCLUSION

In this paper, we presented a 3-tier decentralized data marketplace architectural design with smart contract for managing the terms of the agreement in a way that involves no intermediary. We also outlined the participant's roles and interaction among them.

In this ongoing work, there is no trusted party involved. As a result, there are potential domain threats such as payment fairness, authentication/privacy of the participants, faithful delivery of data and probable malicious behaviour of trust less nodes which needs to be considered while designing it further. As a future work, we planned for a full-fledged implementation to evaluate the performance. A formal security analysis will also be performed.

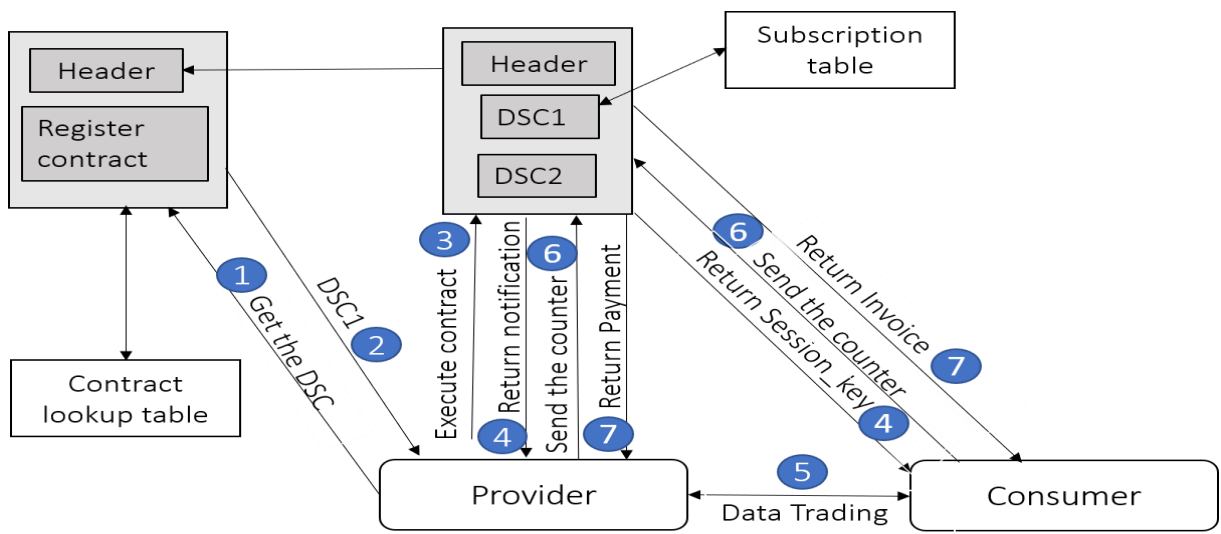

Fig. 8. Smart-contract based data trading.